\documentclass[a4paper,12pt]{article}
\usepackage[utf8]{inputenc}
\usepackage{pythontex}
\usepackage{amsmath}
\usepackage{algpseudocode}
\usepackage{algorithm}
\usepackage{physics}
\usepackage[mathscr]{euscript}
\usepackage{hyperref}
\usepackage{blindtext}
\usepackage{mathtools}
\usepackage{graphicx}
\usepackage{hyperref}
\usepackage{enumitem}
\usepackage{appendix}
\usepackage{comment}
\usepackage[usenames,dvipsnames]{color}
\usepackage{placeins}
\usepackage{bbm}

\usepackage{amsfonts, amsmath, amssymb, multirow, longtable, ntheorem, indentfirst, hanging, mathrsfs}

\newtheorem{theorem}{Theorem}
\newtheorem{lemma}{Lemma}

\hypersetup{colorlinks,linkcolor={blue},citecolor={blue},urlcolor={blue}} 
\usepackage[margin=1in]{geometry}
\usepackage{booktabs}
\usepackage{multirow}
\usepackage{float}
\usepackage{setspace}
\restylefloat{table}
\usepackage{siunitx} 
\usepackage{tikz} 
\usetikzlibrary{shapes.geometric, arrows, positioning}
\tikzstyle{startstop} = [rectangle, rounded corners, minimum width=4cm, minimum height=1cm, text centered, draw=black, text width=4cm]
\tikzstyle{startstopnoborder} = [minimum width=3cm, minimum height=1cm, text centered, text width=3cm]
\tikzstyle{decision} = [diamond, minimum width=1cm, minimum height=0.5cm, text centered, draw=black, text width=3cm,aspect=2]
\tikzstyle{arrow} = [thick,->,>=stealth]

\title{Monitoring of Drift Patterns in Image Data}
\author{Subhasish Basak\footnote{Indian Statistical Institute, Kolkata}, Anik Roy\footnote{Rollins School of Public Health, Emory University, USA}, Partha Sarathi Mukherjee$^\ast$}
\date{\today}

\begin{document}
\maketitle

\def\spacingset#1{\renewcommand{\baselinestretch}%
{#1}\small\normalsize} \spacingset{1}

\begin{abstract}
Sequential monitoring of images has broad applications across various domains, including climate science, ecosystem monitoring, medical diagnostics, and so forth. In many such applications, images acquired over time exhibit gradual changes, referred to as drifts, which pose significant challenges for monitoring. Rather than detecting only abrupt step changes, it is crucial to monitor and characterize these drift patterns. Despite its practical importance, the problem of drift monitoring in image sequences has received limited attention. This paper addresses this gap by proposing a novel drift monitoring method based on an oblique-axis regression tree. It is particularly effective for monitoring drift patterns in the jump location curves present in the image intensity functions. By leveraging a decision tree framework, the method captures discontinuities both in spatial image intensity and temporal progression. A key advantage of this method lies in its flexibility: in the absence of drift, it remains capable of detecting abrupt step changes. Theoretical properties and numerical performance in diverse types of simulation settings indicate its broad applicability.
\end{abstract}

\vspace{0.1 in}



\spacingset{1.1}
\section{Introduction} \label{oblm::intro}
With the widespread availability of advanced image acquisition technologies, large volumes of image data are now routinely collected across various scientific domains. Consequently, images have become a commonly used data format in numerous sectors. A prominent example of such a data source is the Landsat project, jointly led by the U.S. Geological Survey (USGS) and NASA. Since its inception in 1972, this program has launched eight satellites to continuously capture scientifically valuable images of the Earth's surface. The resulting 53-year archive of Landsat images has emerged as a crucial resource for a wide range of scientific disciplines, including forest science, climate science, agricultural forecasting, ecological and ecosystem monitoring, water resource management, biodiversity conservation, and many others. The current Landsat satellite is capable of capturing images of the same geographic region approximately every 16 days. This regular acquisition makes it essential to assess whether the observed images are temporally stable, that is, to sequentially monitor the images for potential temporal changes. Importantly, the need for such monitoring is not limited to satellite imagery; it also extends to various domains such as industrial quality control and medical diagnostics. Sequential image monitoring poses significant challenges, primarily due to the complex structure of the image intensity function, which often includes discontinuities. Moreover, changes in a sequence of images tend to occur gradually over time, a phenomenon referred to as {\it drift}. In many applications, the goal extends beyond detecting abrupt changes or step shifts, to the more complex task of monitoring and characterizing drift patterns. In the context of change-point detection, modeling drift in sequential image data is considerably more challenging. This paper specifically addresses this issue by proposing a novel method for monitoring drift patterns in sequential images. Despite its practical importance, this problem remains largely underexplored in the existing image monitoring literature. To fill this gap, we introduce an algorithm based on decision trees—more precisely, an oblique-axis regression tree (ORT). The rationale for adopting a decision tree-based approach lies in its capacity to effectively model discontinuities in both the spatial and temporal dimensions. The ORT framework offers the flexibility needed to capture complex edge structures in image intensity functions, and to accurately monitor gradual temporal changes and their patterns.

\subsection{Data description and motivation}\label{oblm::data_description}
The methodology developed in this paper is motivated by a sequence of satellite images capturing the Aral Sea region. The Aral Sea, formerly the third-largest lake in the world, is an endorheic lake situated between Kazakhstan and Uzbekistan. Over the past several decades, it has undergone dramatic shrinkage primarily due to unsustainable irrigation practices. This decline has resulted in severe environmental and socio-economic consequences, including increased salinity, frequent dust storms, and significant alterations in the regional climate. It is widely regarded as one of the most devastating environmental crises globally. In response, several initiatives have been undertaken by Kazakhstan and Uzbekistan to slow down the rate of shrinkage. The sequence of satellite images of the Aral Sea presents a compelling dataset for evaluating potential changes in the rate or pattern of this shrinkage over time. In addition to introducing the proposed methodology, this paper applies the algorithm to analyze satellite images of the Aral Sea from 2013 onward. Beyond this specific case, the issue of water body shrinkage is of broader concern: rapid urbanization, increasing water demand, and high evaporation rates, particularly in arid and semi-arid regions, have led to the gradual depletion of several other lakes and reservoirs worldwide. In such contexts, the proposed algorithm offers a valuable tool for monitoring drift patterns and long-term environmental change.

\begin{figure}[ht!]
    \centering
    \includegraphics[width=\linewidth]{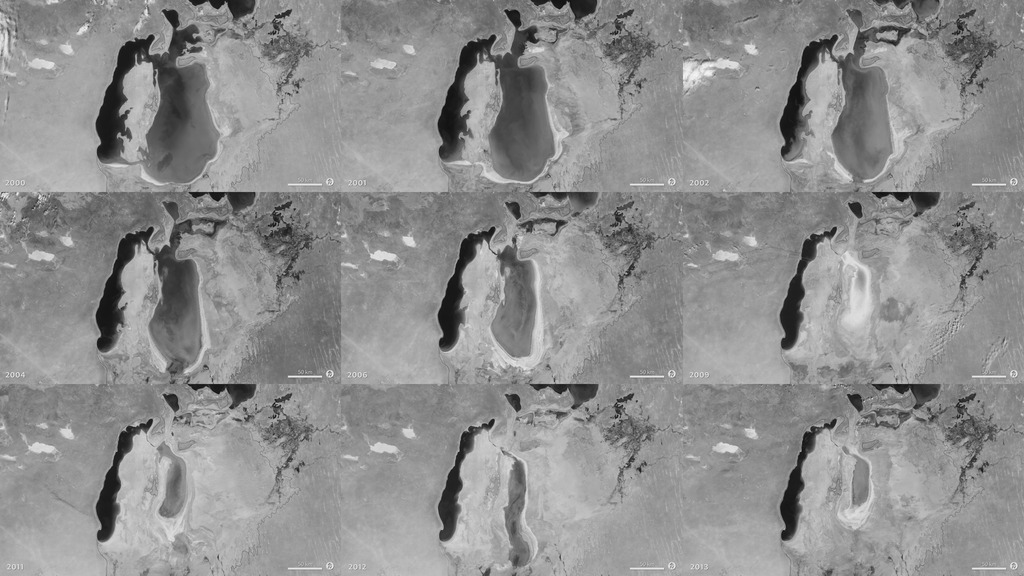}
    \caption{Gradual shrinking of the Aral sea over time. (Images source: NASA). (a) Images at upper panel are captured in 2000, 2001 and, 2002.
    (b)Images at middle panel are captured in 2004, 2006, and, 2009. (c) Images at lower panel are captured in 2011, 2012, and, 2013.}
    \label{oblm::aral_intro}
\end{figure}

The proposed drift monitoring algorithm demonstrates broad applicability across a range of scientific domains. In medical imaging, for example, it can be employed to monitor temporal changes such as tumor progression or regression, thereby supporting the evaluation of treatment efficacy over time. More generally, the methodology holds promise for integration into artificial intelligence frameworks aimed at automated monitoring and decision-making. By enabling the detection and characterization of gradual structural changes and its patterns in complex data, the proposed approach contributes to the advancement of data-driven monitoring systems in both scientific research and industrial applications.

\subsection{Literature review}
Traditional techniques of \textit{statistical process control} (SPC) have been widely used in manufacturing industries to monitor \textit{univariate processes} over time. The main goal of SPC is to observe the process consistently to detect any systematic or non-random variations in the observed data, which is very important for maintaining the quality standards of a product. Recently, there has been a growing interest among researchers in developing control charts specifically designed for monitoring image data. Not only in manufacturing industries, it has diverse applications across various disciplines of science and engineering.  For a related discussion on image monitoring within the framework of SPC methods, see \cite{Qiu2018QE, qiu2020big}. The problem of image surveillance is a challenging problem for the following reasons. First of all, the task of image surveillance is a high volume, high velocity data-rich application. Secondly, the real images often contain noise, and along with noise, the image data itself is complicated in nature. The image intensity surface is not continuous in nature. Therefore, conventional smoothing techniques often fail to accommodate the discontinuous intensity function. Additionally, a typical grayscale image contains a large number of pixels, leading to a high dimensional problem. Therefore, conventional multivariate SPC methods are not appropriate for such applications. Lastly and more importantly, the change in the image sequence is gradual. In this paper, we call this gradual change in the image sequence as \textit{drift}. It is not straightforward to monitor such gradual change or drift in the sequence of images for any change in pattern. In SPC literature, there are methods to monitor drifts for univariate processes.  However, an easy extension of these methods are not possible for the challenges mentioned above. Despite lots of challenges the problem of image surveillance has, several SPC methods have discussed it in the literature. Since, images traditionally been used as a data format in chemical and manufacturing industries, most of the available methods discussed in terms of their applications. A comprehensive review of statistical image monitoring can be found in \cite{megahed2011review}. Prior research in this field is mainly follow the conventional control chart after extracting various features from the images. For instance, \cite{megahed2012spatiotemporal,he2016image} consider a set of region of interest (ROIs) for individual images and use existing GLR (generalized likelihood ratio) and MGLR (multivariate-GLR) control charts, respectively, based on average image intensity at each ROI. For high resolution images, the variance-covariance matrix in the MGLR control chart is not invertible and \cite{okhrin2020new} address this issue by proposing a more sophisticated GLR control chart. The methods proposed by \cite{koosha2017statistical} and \cite{kang2022statistical} follow the profile monitoring approach for images, relying on feature based image monitoring schemes utilizing wavelet coefficients of individual images over time. \cite{roy2025upper} and \cite{yan2017anomaly} discuss anomaly detection in a noisy image sequence with applications in manufacturing industries. \cite{bui2021spc4sts,bui2018monitoring} propose several approaches for monitoring patterns in the images of textured surfaces. In recent years, there has been a growing body of research focused on monitoring high-dimensional tensor data. Examples include methods proposed by \cite{zhen2023image,yang2023tensor,fang2019image}. Additionally, closely related research include image comparison and image monitoring schemes based on the jump location curves (JLCs) in the image intensity functions, e.g.,  \cite{roy2024control, roy2024image, feng2018difference}. Despite extensive literature on image based quality control, the above methods are not primarily designed for drift pattern surveillance. Although there are various control charts for both univariate and multivariate processes in the SPC literature \cite{yi2022adaptive, su2011adaptive}, they are not directly useful for drift pattern monitoring in the image data. The proposed method seeks to address this problem of drift pattern surveillance in images.

\subsection{Novelty and contribution}
In this paper, we introduce a novel image monitoring framework designed to detect changes in the drift pattern within a sequence of multi-temporal images. A shift in the drift pattern refers specifically to a change in the manner in which the jump location curves (JLCs) evolve gradually over time. The proposed method operates as follows: at each time-point, the drift pattern is captured by (i) fitting an oblique-axis regression tree (ORT) to a sequence of preceding images, and (ii) estimating a denoised version of the current image based on the predicted splitting rules derived from the historical sequence. A CUSUM statistic is then constructed by quantifying the overall discrepancy between the denoised image and the observed image at the current time point. The core intuition is that if the current image follows the same drift pattern as the preceding sequence, the denoised image, generated using the inferred splitting rules, should provide a close approximation to the actual image. Consequently, in the absence of a change in the drift pattern, this discrepancy is expected to remain small. The key advantage of the proposed approach lies in its ability to preserve the structural features of the JLCs while effectively tracking their gradual temporal evolution. Although such pattern monitoring of gradual changes in JLCs is prevalent in many practical settings, it has received limited attention in the image monitoring literature. A recent effort by \cite{yi2023water} addresses a related problem in the context of satellite imagery; however, their approach focuses on calculating the area of image objects and employs a univariate likelihood ratio-based framework for drift detection. In contrast, the method proposed in this paper is tailored to handle more general image types and is particularly well-suited for monitoring drift patterns in JLCs, offering a more flexible and structurally informed alternative.

The main contributions of this paper and key highlights of the proposed methodology are summarized below:
\begin{itemize}
    \item The core novelty of the proposed method lies in the introduction of a CUSUM-based framework for detecting changes in the pattern of gradual evolution within a sequence of images.
    
    \item Leveraging the ORT-based construction, the method effectively preserves jump discontinuities in the image intensity surface, which are often critical features in many monitoring applications.
    
    \item By focusing primarily on the evolution of jump location curves (JLCs), the method inherently suppresses insignificant background anomalies which is an especially desirable property in satellite imaging where such anomalies are common.
    
    \item Although the methodology is designed to detect changes in drift patterns, it is also capable of signaling abrupt step shifts in the absence of any drift, thereby offering a high degree of adaptability.

    \item The proposed framework is broadly applicable to a wide range of image monitoring tasks. Its practical utility is demonstrated through extensive simulation studies which highlight its superior performance, and a real-world application.
\end{itemize}

\subsection{Organization}
The remainder of the paper is organized as follows. Section \ref{oblm::methodology} introduces the proposed methodology in detail. Section \ref{oblm::stat_prop} discusses theoretical properties of the proposed control chart. Section \ref{oblm::numericals} presents numerical studies, including both simulation experiments and a real-world application. Finally, a few remarks in Section \ref{oblm::conclusion} conclude this paper.

\section{Proposed Methodology}\label{oblm::methodology}
This paper aims to track the pattern of the drift in the sequence of image data. While the current methodology is primarily designed for a sequence of $2$D grayscale images, it is also applicable to image sequences in other dimensions e.g., sequence of $3$D images. In this section, we present the proposed methodology in a step-by-step manner. 

\subsection{Image pre-processing} The initial step in image monitoring is the image pre-processing, to make the sequence of images in a usable format. In satellite images, often the images are not geometrically aligned. For reliably monitoring the images, they should be geometrically aligned. In the literature, this alignment process is commonly referred to as \textit{image registration} \cite{qiu2013feature,xing2011intensity}. In this paper, we assume that the image sequence is already registered. In case it's not, we apply an image registration algorithm before proceeding with monitoring. Pre-processing also involves tasks such as resizing the images to the same resolution, scaling the image intensities to a given range, or removing irrelevant background regions.

\subsection{Oblique-axis regression tree}
Tree-based methods are widely used in modern statistics and machine learning. In this paper, we develop a monitoring scheme where the control statistic is constructed using an approach based on a special type of decision tree called the oblique-axis regression tree (ORT). The choice of ORT is mainly motivated by discontinuous nature of the image intensity function, which often displays an image object with complex shape. Traditional axis-aligned CART algorithms such as \cite{breiman1984classification} create splits based on individual covariates while the search space is restricted to the standard axis directions. In contrast, oblique CART permits splits based on linear combinations of covariates, thus expanding the search space. The key distinction between standard CART and oblique CART lies in how they partition the predictor space. While the former creates rectangular regions, ORT produces convex polytopes. Since the image intensity functions are often discontinuous and the locations of the discontinuities may not align with the coordinate axes, ORT offers a more suitable approach for accurately capturing the object boundaries or the edges of the images. For a comprehensive discussion on ORT, readers are referred to \cite{zhan2024consistencyobliquedecisiontree,cattaneo2024convergence}. A recent research focusing on the effectiveness of ORT in preserving jumps in discontinuous regression surfaces can be found in \cite{basak2025estimation}.

\subsection{Jump regression on a sequence of images}
\noindent Under jump regression analysis (JRA) \cite{qiu2005image}, a sequence of 2D $n\times n$ grayscale images can be considered to follow the model:
\begin{equation}\label{oblm::eq:image_seq_model}
    \mathcal{I}_k(i,j)=w(x_i,y_j,t_k)= f(x_i,y_j,t_k)+\varepsilon(x_i,y_j,t_k) \ \ \text{for} \ \ i,j=1,2, \ldots, n; \ k= 1,2, \ldots, 
\end{equation}
where $\mathcal{I}_k$ is the observed image at $k$-th time point, $w(x_i,y_j,t_k)$ is the observed image intensity at $(i,j)$-th pixel coordinate at $k$-th time point $t_k$, $f(x_i,y_j,t_k)$ is the corresponding unknown true image intensity value and $\varepsilon(x_i,y_j,t_k)$ are the independent and identically distributed (i.i.d.) pointwise random noise with mean $0$ and variance $\sigma^2$. For the image data, the pixel coordinates lie on an equally spaced lattice defined on the design space $\Omega=[0,1] \times [0,1]$, namely, $\{(x_i,y_j)= (i/n,j/n): i,j=1,2,\ldots, n\}$. The observed time points are also assumed to be equally spaced. Under the JRA model, the true image intensity function is continuous in the design space $\Omega$ for each $t_k$, except on the edges in the boundary of the image object. In this paper, we assume that $f$ is piecewise constant for simplicity. Therefore, $f$ can also be written as:
$$f(x,y,t)=\sum\limits_{\ell=1}^{J} c_\ell \mathbbm{1}_{\Gamma_\ell}(x,y,t),$$
where $\{\Gamma_\ell: \ell = 1, 2, \ldots, J\}$ is a partition of $\Omega$. The boundaries of these regions are the discontinuity points of the image. These are also called jump location curves or JLCs in the JRA literature. In this paper, we also assume that the JLCs are piecewise linear except for a few finitely many points. For ease of notations, we will be assuming that JLCs are linear, i.e, we will be considering each linear part of the piecewise linear JLCs to be different JLCs. In this context, we consider the sequence of image as a three dimensional image, so any change or discontinuity along the time dimension will also be reflected in the JLCs.

\noindent \textit{Definition 1:} We say that the point $(x_i,y_j)$ is drifting in direction $(\beta_1,\beta_2)$ if: 
\[f(x_i+\beta_1*t,y_j+\beta_2*t,t)=f(x_i,y_j,0).\]

From the above definition, some confusions arise because the image intensity function is locally constant in certain regions. Moreover, the above definition does not emphasize the fact that all points in a local neighborhood should drift in the same direction. Therefore, it is more appropriate to discuss drifts on image objects instead of points. As JLCs usually surround an object, we now discuss drifts in a sequence of images in term of the drifts of JLCs.

\noindent \textit{Definition 2:} Suppose that $\mathcal{I}_k$ is the observed image captured at time $t_k$, and $\Gamma$ is a JLC on it, then we say that $\Gamma$ is drifting if all points on it are drifting at the same direction.

Any linear JLC in $\mathcal{I}_{k}$ can be expressed in the form $\alpha_1 x + \alpha_2 y <c, (x,y)\in S$ where S is a subset of $\Omega$. If it is drifting, then we can rewrite this as $\alpha_1 (x+\beta_1t) + \alpha_2 (y+\beta_2t) <c \implies \alpha_1 x + \alpha_2 y + \alpha_3 t <c$, where $\alpha_3 = \alpha_1\beta_1+\alpha_2\beta_2$. If $(\beta_1,\beta_2)$ changes, say to $(\beta^\prime_1,\beta^\prime_2)$, then $\alpha^\prime_3=\alpha_3 \implies \alpha_1\beta_1+\alpha_2\beta_2=\alpha_1\beta^\prime_1+\alpha_2\beta^\prime_2 \implies  \frac{\beta^\prime_1-\beta_1}{\beta^\prime_2 - \beta_2}=-\frac{\alpha_2}{\alpha1}$. Therefore, if this relationship does not hold, it means that $\alpha_3$ has changed.

Now this situation implies that the JLC is moving along the resultant of the previous direction and a direction parallel to the JLC itself. So for piecewise constant images it should be visually indistinguishable from the starting image unless there is another JLC present in the image which is not parallel to the first JLC. So changes of this nature, will be reflected in the drift of some other JLC.

In the oblique regression tree method, since we split according to similar rules, it should be able capture drifting JLCs properly, and it should only detect a change with respect to time if the slope $(\alpha_1,\alpha_2,\alpha_3)$ changes abruptly at some point. One thing to note here that is that the proposed method cannot determine $(\beta_1,\beta_2)$ exactly, although we will be able to detect any change in that direction.

\subsection{Construction of the ORT based decision tree}
\label{oblm::ssec::tree}
 In this subsection, we discuss the recursive tree splitting algorithm, a pivotal step for the proposed drift pattern monitoring method.  For any given node $\mathcal{N}$, direction $\alpha$ and constant $c$, we define $\mathcal{N}_L= \{{z}:{\alpha}^T{z} \leq c\}$ and $\mathcal{N}_R=\{{z}:{\alpha}^T{z} > c\}$ as the children nodes, where $z = (x,y,t)$. The key idea behind the tree construction is to hierarchically partition the data by greedy binary splitting algorithm. Now for a given parent node $\mathcal{N}$ and the splitting rule denoted by the pair $(\alpha,c)$, we define:
\begin{equation}\label{oblm::eq:impurity_gain}
    \widehat{\Delta}(\mathcal{N},{\alpha},c)= \frac{1}{|\mathcal{N}|}\bigg[\sum\limits_{{z} \in \mathcal{N}}\big(w({z})-\Bar{w}\big)^2 - \sum\limits_{{z} \in \mathcal{N}_L}\big(w({z})-\Bar{w}_L\big)^2-\sum\limits_{{z} \in \mathcal{N}_R}\big(w({z})-\Bar{w}_R\big)^2 \bigg],
\end{equation}
where $\bar{w}$,$\Bar{w}_L$, and $\Bar{w}_r$ are the average image intensities over the respective nodes $\mathcal{N}$, $\mathcal{N}_L$, and $\mathcal{N}_R$. Note that $\widehat{\Delta}(\mathcal{N},{\alpha},c)$ is the reduction in sum of square (SSE) if we divide the node using the rule  $(\alpha,c)$. We denote the maximum value of $\widehat{\Delta}(\mathcal{N},{\alpha},c)$ over $(\alpha,c)$ by $\widehat{\Delta}(\mathcal{N})$. We split the node $\mathcal{N}$ if the value of $\widehat{\Delta}(\mathcal{N})$ is greater than a pre-determined threshold. Now, to construct the regression tree, we start by considering all observed intensity values across the images and time as a node, label them with their co-ordinates respectively, and then follow Algorithm \ref{oblm::algo_splitting}.

\begin{algorithm}\label{oblm::alg::tree}
\begin{algorithmic}[H]
\caption{Recursive Tree Construction}

\vspace{2mm}

 \State \textbf{Input:} A sequence of images labeled with respect to time.
 \State \textbf{Start:} We have one node $\mathcal{N}$, the root, in the tree.
 \State Calculate $\widehat{\Delta} (\mathcal{N}).$ 
 \If{$\widehat{\Delta} (\mathcal{N}) \leq \mbox{cutoff}$}
    \State Stop and return the tree you already have.
\Else
    \State Calculate $(\widehat{{\alpha}}_{opt},\widehat{c}_{opt})=  \arg \max \limits_{({\alpha},c)}\widehat{\Delta}(\mathcal{N},{\alpha},c)$.
    \State Split $\mathcal{N}$ into $\mathcal{N}_R$ and $\mathcal{N}_L$ using the line $\{{\widehat{\alpha}^T_{opt}}{z} = \widehat{c}_{opt}\}$.
\EndIf
\State Calculate trees $T_1$ and $T_2$ recursively, using $\mathcal{N}_R$ and $\mathcal{N}_L$ as root nodes.
\State Join trees $T_1$ and $T_2$ with $\mathcal{N}$ to get the final tree $T$.
\State \textbf{Output:} $T$ is the estimated tree.
\label{oblm::algo_splitting}
\vspace{2mm}

\end{algorithmic}
\end{algorithm}

Consider a sequence of images of length $m$ with labeled time points, denoted as $\{\mathcal{I}_1,\mathcal{I}_2, \ldots, \mathcal{I}_m\}$ following the jump regression model expressed in Eqn. \eqref{oblm::eq:image_seq_model}.
Now that after implementing the binary splitting algorithm we have the estimated tree $T$, consider the corresponding set of decision rules denoted as $T_{rule}$, which consists of all decision rules while constructing the tree. Suppose, we are given an image $\mathcal{I} \notin \{\mathcal{I}_1,\mathcal{I}_2, \ldots, \mathcal{I}_m\}$ labeled with time $t_{\mathcal{I}}$. Then, by applying the rule $T_{rule}$ on that image, we get:
\[
T_{rule}(\mathcal{I},t_{\mathcal{I}}) = \{\mathcal{L}_{T,\mathcal{I}}^i|i \in \{1,2,\dots,K_n^m(\mathcal{I}) \}\},
\]
where $K_n^m(\mathcal{I})$ is the number of leaf nodes that depends on the image, its resolution and the length of the image sequence, and $\mathcal{L}_{T,\mathcal{I}}^i$ are the leaf nodes in the image $\mathcal{I}$ after applying the decision rules $T_{rule}$. Typically, for a given sequence, $m$ and $n$ are kept fixed in the proposed algorithm. Therefore, to make the notation less cumbersome, we will denote it with $K(\mathcal{I})$ later in the paper. Intuitively, the intensity values in a leaf node only contains those from similar pixel coordinates. Then, the proposed denoised estimator of $t_{\mathcal{I}}$ at the point $(x,y)$ is based on the leaf-only averaging, which is given by
\begin{equation}
\label{oblm::eq:esti}
    \widehat{f}_{t_{\mathcal{I}}}(x,y) = \frac{1}{|\mathcal{L}_{T,\mathcal{I}}(x,y,t_{\mathcal{I}})|}\sum\limits_{(x_p,y_q,t_I) \in \mathcal{L}_{T,\mathcal{I}}(x,y,t_I)} \mathcal{I}(p,q),     
\end{equation}
where $\mathcal{I}(p,q)$ denotes the intensity of the image at the pixel $(x_p,y_q)$ and $\mathcal{L}_{T,\mathcal{I}}(x,y,t_\mathcal{I})$ corresponds to that leaf node in $T_{rule}(\mathcal{I}, t_{\mathcal{I}})$ which contains the point $(x,y,t_\mathcal{I})$.

\vspace{0.25cm}
\noindent\textit{Remarks:} Note that the above estimator based on the leaf-averaging is a jump preserving estimator of the image intensity function accommodating the temporal pattern or the drift. If $\mathcal{I} \in \{\mathcal{I}_1,\mathcal{I}_2, \ldots, \mathcal{I}_m\}$, then the estimated image is the denoised image based on the drift information provided by the known sequence. Hence, this method can also be used as an image denoising method for the sequence of images. The proposed image monitoring method under a drift can, therefore, accommodate the jumps in the discontinuous image intensity function. For certain applications, to accommodate the intricate details of the image, the leaf-only-averaging may not work well. In those applications, one can modify the proposed method by introducing local-leaf-only averaging. Theoretical consistency of the proposed algorithm can be shown similarly as provided in \cite{basak2025estimation}.

 \subsection{Phase II monitoring using ORT}   
In traditional SPC literature, process monitoring is typically divided into two phases: Phase I and Phase II. In this paper, we propose a CUSUM control chart for Phase II monitoring of the temporal drift pattern in the image sequence based on the ORT framework. The proposed approach consists of two major components: Firstly, building a predictive model that captures the drift pattern; and secondly, formulating a CUSUM statistic to detect changes in the drift pattern. In Phase II stage, the observed image at the $k$-th time-point can be expressed by following the same model as given in Eqn. \eqref{oblm::eq:image_seq_model}. To enable online monitoring of the sequential process, we assume that an initial in-control (IC) dataset $\Psi_{M, IC}^{(0)}=\{\mathcal{I}_{-M+1},\mathcal{I}_{-M+2},\ldots , \mathcal{I}_{0}\}$ of size $M$ to be available. As the change occurs gradually, such prior data are required to capture the underlying drift pattern, and to start the online monitoring. The time period of the initial IC dataset is then set as a baseline time interval. The primary objective of online monitoring is to detect any substantial deviation in the future drift pattern of the process from its regular drift pattern occurring in the baseline time interval as promptly as possible. Now at any time point $t_{k-1}$, a regression tree is fitted using the available image data from the past. Since involving all prior images is computationally challenging, we adopt a moving window approach of length $m_0$, and fit the ORT on the most recent $m_0$ number of images only.  In this paper, we refer to $m_0$ as the window width. We choose the window width $m_0$ to be strictly smaller than the total number of available observations in the baseline time interval so that the remaining $\left(M-m_0\right)$ images can be used to estimate the average prediction error, acceptable drift, and control limits, and so forth when these values are unknown. Additionally, it is a natural assumption that the choice of $m_0$ should cover the entire cycle of a full drift pattern. Now at time $t_{k-1}$, we have the partition based on the regression tree with images from time $t_{(k-m_0)}$ to $t_{k-1}$, and we use this to predict the partition of the image captured at time $t_{k}$. To this end, we compute the denoised image at the time-point $t_k$ based on the predicted tree partition. The predicted partition can be calculated by putting $t=t_k$ at the constructed decision rule based on the previous $m_0$ number of images. Mathematically, based on the image sequence $\Psi^{(k-1)}_{m_0}=\{\mathcal{I}_{(k-m_0)},\ldots, \mathcal{I}_{(k-1)} \}$, consider the data with dimension $n\times n\times m_0$, and fit a decision tree on it using the greedy algorithm mentioned in Section \ref{oblm::ssec::tree}. Then, similar to Eqn. \eqref{oblm::eq:esti}, the denoised image at the $k$-th time point with the predicted partition based on the image sequence $\Psi^{(k-1)}_{m_0}$ is defined as 
\[
    \widehat{f}_{t_k}^P(x,y) = \frac{1}{|\mathcal{L}_{T,k}^P(x,y,t_k)|}\sum\limits_{(i,j,t_k) \in \mathcal{L}_{T,k}(p,q,t_k)} \mathcal{I}_{t_k}(p,q), 
\]
where $\mathcal{L}_{T,k}^P(x,y,t_k)$ corresponds to the leaf node at the $k$-th time point based on the predicted decision rule by putting $t=t_k$ in the constructed decision rule on the image sequence $\Psi^{(k-1)}_{m_0}$. $\mathcal{I}_{t_k}(p,q)$ indicates the observed image intensity at $(x_p,y_q)$-th coordinate of image captured at the $k$-th time-point, i.e., $\mathcal{I}_k$. Then, the overall difference between the observed image at $t_{k}$ and predicted denoised image at $t_{k}$ is defined as 
\begin{equation}\label{oblm:eq:overall_diff}
    \Lambda_{t_k}=\frac{1}{(n^2-K(\mathcal{I}_{t_k}))\theta^2} \sum_{i,j}^n (\widehat{f}_{t_k}^P(x_i,y_j)-w(x_i,y_j,t_k))^2,
\end{equation}
where $\theta^2$ is the sum of noise variance and squared prediction error. In practice, we estimate $\theta^2$ by the average MSE of observed in-control data when compared to the estimates given by the ORT algorithm. A mathematical formulation of this is given in Theorem \ref{oblm::mainthm}.
\begin{algorithm}[b!]
    \begin{algorithmic}[H]
    \caption{The proposed drift pattern monitoring scheme for a sequence of images}
    \State \textbf{Input:} A sequence of $n \cross n$ images labeled with respect to time, available at the $k$-th time-point, window width $m_0$ for computation of estimates, value for acceptable amount of change $\kappa$, an estimated value of noise $\theta^2$, and a cutoff value.\\
     \State \textbf{Start:} We start with $k=1$.\\
     \For{$k<\textit{(the length of the image sequence)}$}\\
     \State Fit a tree using algorithm \ref{oblm::alg::tree} using images in $\Psi^{(k-1)}_{m_0}$.\\ 
     \State Calculate $T_{rule}^{t_k}$, the decision rules at time $t_k$, based on the tree obtained.\\
     \State Using $T_{rule}^{t_k}$, calculate $\widehat{f}_{t_k}(x,y)$.\\    
     \State Calculate $\Lambda_{t_k}$ and $Q_{t_k}$.\\
     \State Report $Q_{t_k}$.\\
     \State $k=k+1$\\
     \EndFor
     \label{oblm::algo_monitoring_scheme}
    \end{algorithmic}
\end{algorithm}
Observe that, the denominator in Eqn. \eqref{oblm:eq:overall_diff} is the degree of freedom of the sum, as there are only $K(\mathcal{I}_{t_k})$ non-empty leaf-nodes in the fitted tree. Additionally, it is important to note that, if there is no change in the drift pattern of the image, this estimate should be close enough to the observed image and $\Lambda_{t_k}$ would be small because it contains only the noise. Conversely, if a change in the drift pattern has occurred, then the predicted denoised image and the observed image should not be similar and $\Lambda_{t_k}$ would be large. Therefore, by calculating the prediction error, we can detect if a change in the drift pattern has occurred between the time-points $t_{k}$ and $t_{(k+1)}$. Theorem \ref{oblm::mainthm} shows that after suitable centering and scaling, the mean squared error asymptotically follows a normal distribution. Thus, a natural CUSUM charting
statistic can be defined by
\begin{equation} \label{cusum}
    Q_{t_k}=max\bigg(0, Q_{t_{(k-1)}}+\frac{n}{\sqrt{2}}\left(\Lambda_{t_k}-1\right)-\kappa\bigg), \quad k=1,2, \ldots,
\end{equation}
where $Q_{t_0}=0,$ and $\kappa \geq 0$ is a pre-specified allowance parameter. To choose the allowance parameter optimally, the readers are referred to \cite{qiu2013introduction}. The proposed CUSUM statistic, raises a signal if 
$$Q_{t_k} > q_0,$$ where $q_0$ is a pre-determined control limit. Traditionally, the control limit is chosen in such a way that the in control {\it average run length} (ARL) of the process can reach the pre-fixed nominal level of $ARL_0$. The in-control ARL is defined as the expected time
to signal when there is no change in the drift pattern.


\begin{figure}[ht!]
\centering
\resizebox{12cm}{!}{
\begin{tikzpicture}[node distance=4cm]
\node (control) [startstop] {\textbf{Start with in-control data $\Psi^{(0)}_{M, IC}$.}};
\node (proc) [startstop,left of=control,xshift=-2cm ] {Perform pre-processing of the images.};
\node (new) [startstop,below of=proc,yshift=1cm] {Collect the new image $\mathcal{I}_{t_k}$ at the $k$-th time point for online monitoring. Pre-process, if necessary.};
\node (rule) [startstop,below of=control,yshift=1cm] {Construct the set of decision rules based on the last $m_0$ images.};
\node (stat) [startstop,below of=rule,xshift=3cm,yshift=1cm] {Calculate the CUSUM charting statistic as in Eqn. \eqref{cusum}.};
\node (choice) [startstop,left of =stat,xshift=-2cm] {Is the CUSUM statistic $>$ threshold?};
\node (iter) [startstop,left of =choice,xshift=-2cm] {Set $k=k+1$.};
\node (fin) [startstop,below of=choice,yshift=1cm]{\textbf{Raise a signal for pattern change.}};
\draw [arrow] (control) -- (proc);
\draw [arrow] (proc) -- (new);
\draw [arrow] (new) -- (rule);
\draw [arrow] (rule) -| (stat);
\draw [arrow] (stat) -- (choice);
\draw [arrow] (choice) -- node[anchor=south]{No}(iter);
\draw [arrow] (iter) |- (new);
\draw [arrow] (choice) -- node[anchor=west]{Yes} (fin);
\end{tikzpicture}
}
\caption{The flowchart diagram for the proposed method for monitoring drift pattern in an image sequence.}
\label{chap::oblm::fig::algo}
\end{figure}
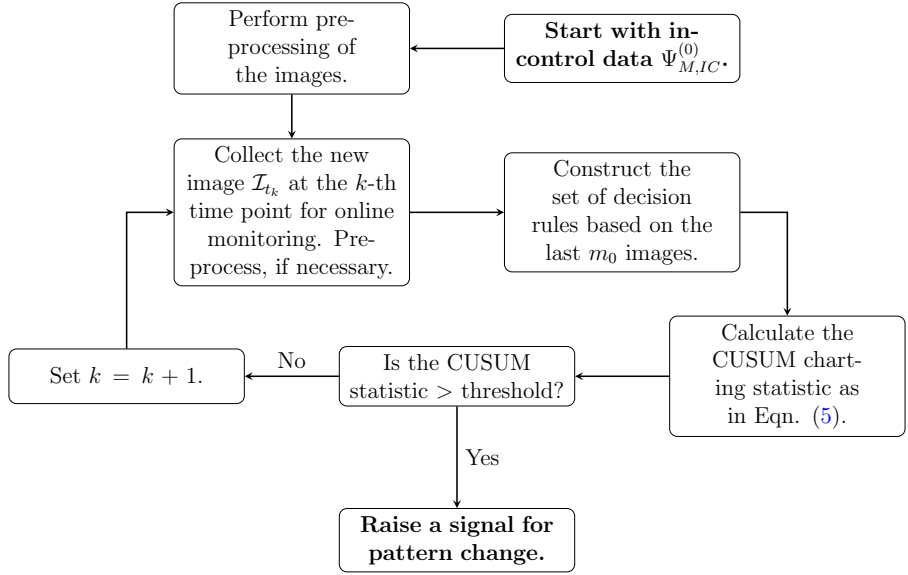

\section{Statistical Properties}\label{oblm::stat_prop}
In this section, we investigate certain theoretical properties of the proposed the online drift detection algorithm. Before going into the main results, we introduce the regulatory assumptions below which are reasonable in a wide range of practical applications.
\begin{enumerate}[label = (\roman*)]
    \item $f$ is piecewise Lipschitz.
    \item $f$ is bounded inside its domain $[0,1]^2$.
    \item The points in $\overline{\Gamma_\ell}\backslash int(\Gamma_\ell)$ are called jump points, and the set of jump points have Lebesgue measure zero.
    \item There exists at most finitely many JLCs.
    \item Each JLCs is a line segment.
    \item If $f$ is continuous at a jump point, then that point is called a singular point.
    \item The point-wise noise $\varepsilon$ are independently distributed and distributed as $\mathbf{N}(0,\sigma^2)$.
\end{enumerate}

Suppose that we are given a sequence of images of resolution $n \times n$ captured at time-points $\{t_1,t_2,\dots\}$. Consider the following test of hypothesis:
\begin{center}
    $\mathcal{H}_0:  \textit { Drift pattern of the JLCs have not changed}$\\
    $ \textit{vs.}$\\
    $\mathcal{H}_1:   \textit{ Drift pattern of at least one JLC has changed}$
\end{center}
The following Theorem provides asymptotic behavior of the charting statistic under both $\mathcal{H}_0$ and $\mathcal{H}_1$.
\begin{theorem}
Suppose that $\sup_i|t_{i+1}-t_i|\rightarrow 0$ as $n\rightarrow\infty$, and the assumptions listed above hold true. Then, we have the following results:

\noindent (i) Under $\mathcal{H}_0$, $\frac{n}{\sqrt{2}}(\Lambda_{t_k}-1) \overset{d}{\rightarrow} \mathcal{N}(0,1)$.\\
(ii) Under $\mathcal{H}_1$, $\frac{n}{\sqrt{2}}\Big(\Lambda_{t_k}-1-O(1)\Delta^2\Big) \succ \mathcal{N}\left(0,1+\frac{E_{t_k}+O(1)\Delta^2}{1+E_{t_k}}\right)$, where $\Delta > 0$ is the deviation associated with the change in the drift pattern, and $E_{t_k}$ corresponds to the estimation error related to ORT based image smoothing.
    \label{oblm::mainthm}
\end{theorem} 
The above Theorem is a direct consequence of the results provide by \cite{basak2025estimation}. A sketch of the proof is provided in the Supplementary Materials which are attached at the end of this paper.

\section{Numerical Studies}\label{oblm::numericals}
In this section, we first check the performance of the proposed drift monitoring method on various types of simulated data. Next, we compare these performances with those of various state-of-the-art image monitoring methods. Subsequently, we apply the proposed method on a real satellite data on the Aral sea.

\subsection{Performance of the proposed method on various types of simulated data}\label{oblm::simulated}
We conduct extensive simulation studies to evaluate the performance of the proposed algorithm under various types of drift conditions. For demonstration, the images are simulated with JLCs consisting only of straight line segments, and we consider various types of drift along these segments to mimic real-world scenarios. Moreover, we add Gaussian noise with a fixed standard deviation of $0.15$ to each of the simulated images having resolution $128 \times 128$. We set the window width $m_0$ to be $20$ in all simulation examples.

\begin{figure}[ht!]
    \centering
    \includegraphics[width=\linewidth]{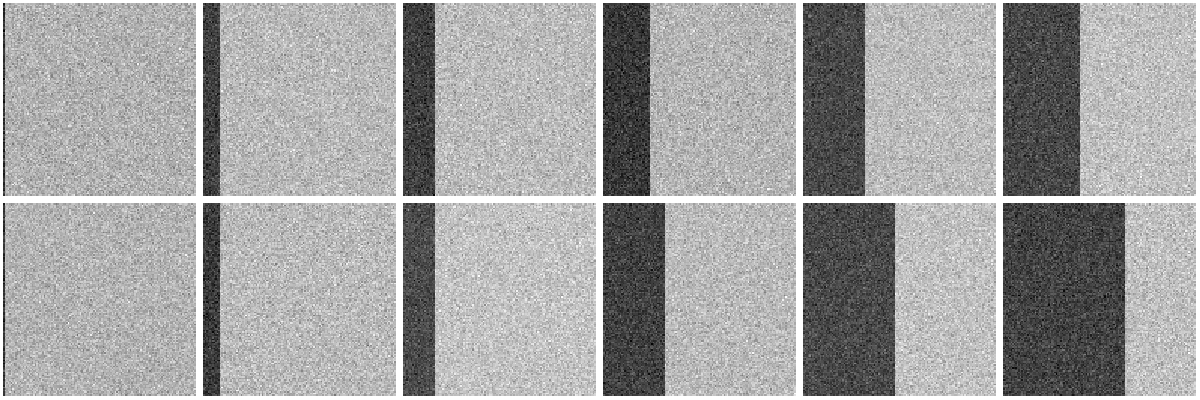}
    \caption{The top row shows the in-control images at time-points $t=$ $0$, $10$, $20$, $30$, $40$, and $50$ for Simulation 1. The bottom row shows the corresponding images for the out-of-control image sequence.}
    \label{oblm::sim1}
\end{figure}

\textbf{Simulation 1:} In the first image sequence, we consider one single JLC consisting of a vertical straight line moving horizontally as the reference. In this sequence, the line segment shifts one pixel to the right per unit of time. Figure \ref{oblm::sim1} illustrates how the image changes over time. These changes occur linearly or at the same rate in the in-control image sequence. We try various values of the allowance parameter $\kappa$, and find the corresponding control limit $q_0$ such that the in-control ARL becomes $20$ based on $100$ independent replications for each trial value of $\kappa$. We then select that value of $\kappa$ and the corresponding $q_0$ for which the plot of the CUSUM statistic visually exhibits minimal trend. In this case, the selected value of $\kappa$ is $0.9$. Using that allowance parameter we determine the control limit $q_0$ for achieving the average run length (ARL) of $20$ indicating that the test of hypothesis has a significance level of $0.05$. Next, we generate a sequence where the drift rate of the line segment doubles from a pre-determined time of $t=20$ in this example. We then calculate the detection delay using the same cutoff and permissible change values. The results show that the proposed algorithm can successfully detect the drift change, with the average detection delay being very close to $1$. The functional form of these two image sequences without noise are as follow:
\begin{align*}
    f_{IC}(x,y,t) &= \mathbbm{1}\Big[x > \frac{t}{128}\Big], \\
    f_{OC}(x,y,t) &= \mathbbm{1}\Big[x > \frac{t + t\mathbbm{1}(t >20)}{128}\Big],
\end{align*}
where $f_{IC}$ and $f_{OC}$ represent the in-control and the out-of-control image sequence, respectively.

\begin{figure}[ht!]
    \centering
    \includegraphics[width=\linewidth]{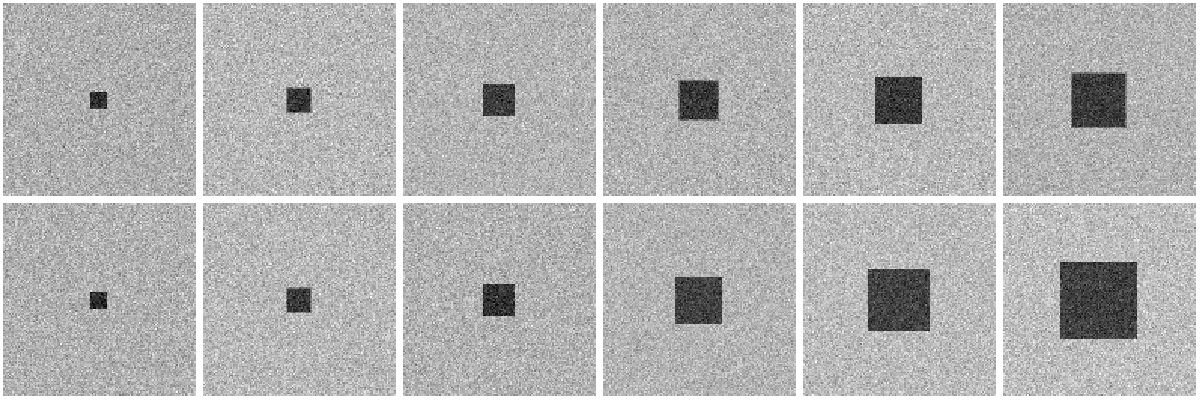}
    \caption{The top row shows the in-control images at time-points $t=$ $0$, $10$, $20$, $30$, $40$, and $50$ for Simulation 2. The bottom row shows the corresponding images for the out-of-control image sequence.}
    \label{oblm::sim2}
\end{figure}

\textbf{Simulation 2:} The second image sequence features a square with four linear JLCs as its arms. In this sequence, the JLCs move away from the center of the square at a rate of $1/4$-th of a pixel length per unit of time. This sequence is designed to simulate a change in size or scale of an object. Figure \ref{oblm::sim2} illustrates how the image changes over time. Following a similar procedure as in the previous simulation, we select the allowance parameter $\kappa$ to be $2.0$, and the corresponding control limit $q_0$ so that the in-control ARL becomes $20$. Similar to the previous case, we double the rate of change after the predetermined time of $t=20$, resulting in an image sequence where a change in the drift pattern occurs. Using the obtained control limit, we find that the average detection delay in this case is close to $4$. Note that in this sequence, it takes $4$ units of time to observe a change in the image for the in-control case, and since the standard deviation of the added noise is relatively large, this result is expected. The functional forms of the image sequences without noise are as follow:
\begin{align*}
    f_{IC}(x,y,t) &= \mathbbm{1}\Big[\max \big\{|x-0.5|,|y-0.5|\big\} > \frac{t+20}{512}\Big],\\
    f_{OC}(x,y,t) &= \mathbbm{1}\Big[\max \big\{|x-0.5|,|y-0.5|\big\} > \frac{20+t+ t\mathbbm{1}(t >20)}{512}\Big].
\end{align*}

\textbf{Simulation 3:} In this sequence, we analyze a square moving from the left to the right of the image frame at a speed of $1/2$ of a pixel length per unit time. Following similar procedure as in the previous two simulation cases, we find that the detection delay to be approximately $2$. This result is consistent with the findings from the second example. In this case, $\kappa$ is selected to be $0.7$. The functional form of these image sequences are as follow:
\begin{align*}
    f_{IC}(x,y,t) &= \mathbbm{1}\big(max(|x-\frac{t}{256})|,|y-0.5|)>\frac{5}{128}\big),\\
    f_{OC}(x,y,t) &= \mathbbm{1}\big(max(|x-\frac{t+ t\mathbbm{1}(t >20)}{256}|,|y-0.5|)>\frac{5}{128}\big).
\end{align*}

\begin{figure}[ht!]
    \centering
    \includegraphics[width=\linewidth]{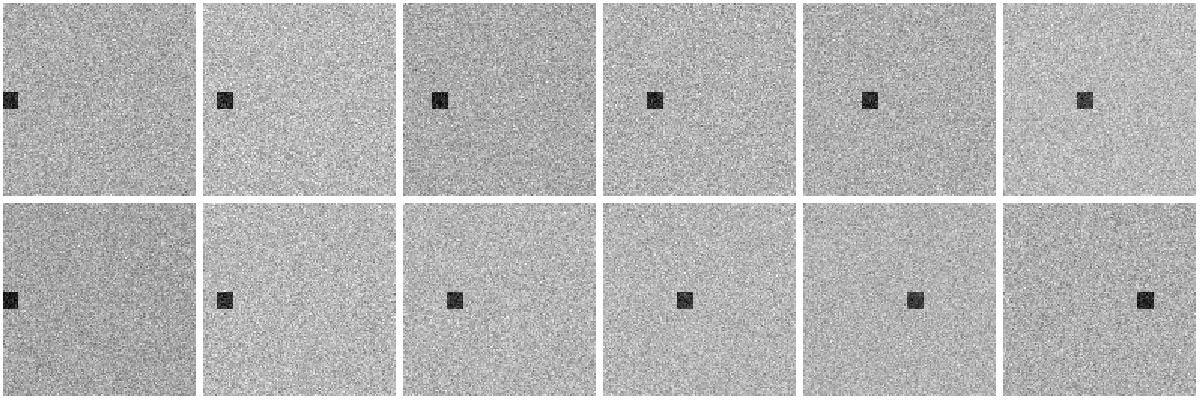}
    \caption{The top row shows the in-control images at time-points $t=$ $0$, $20$, $40$, $60$, $80$, and $100$ for Simulation 3. The bottom row shows the corresponding images for the out-of-control image sequence.}
    \label{oblm::sim3}
\end{figure}

\textbf{Simulation 4:} In the fourth simulation example, we demonstrate that the proposed method performs well when there is no drift in the in-control image sequence, and there occurs a change in JLC in the out-of-control image sequence. Following similar procedure as in the previous simulation cases, we find that the average detection delay is close to $1$, indicating that the proposed method can also be used to monitor images for any change in JLCs when there is no drift. In this case, $\kappa$ is selected to be $2.0$. The following two functions represent the in-control and out-of-control image sequences:
\begin{align*}
    f_{IC}(x,y,t) &= \mathbbm{1}\bigg((x,y)\notin \big[0.25,0.75\big]\cross \big[0.25,0.75\big]\bigg), \\
    f_{OC}(x,y,t) &=\mathbbm{1}(t\leq20)\mathbbm{1}\bigg((x,y)\notin \big[0.25,0.75\big]\cross \big[0.25,0.75\big]\bigg) \\
    &+\mathbbm{1}(t>20)\mathbbm{1}\bigg((x,y)\notin \big[0.25-\frac{5}{128},0.75+\frac{5}{128}\big]\cross \big[0.25-\frac{5}{128},0.75+\frac{5}{128}\big]\bigg).
\end{align*}

\begin{figure}[ht!]
    \centering
    \includegraphics[width=\linewidth]{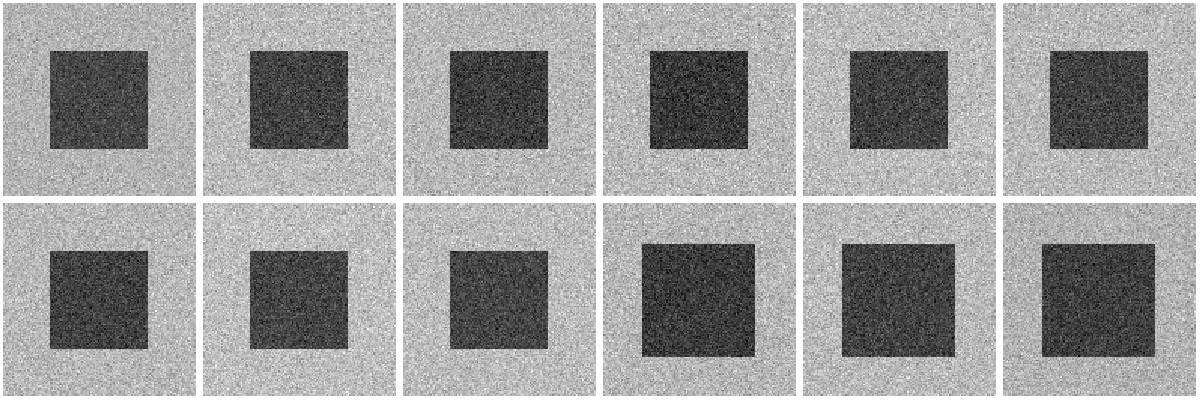}
    \caption{The top row shows the in-control images at time-points $t=$ $0$, $10$, $20$, $30$, $40$, and $50$ for Simulation 4. The bottom row shows the corresponding images for the out-of-control image sequence.}
    \label{oblm::sim5}
\end{figure}

\begin{figure}[ht!]
    \centering
    \includegraphics[width=\linewidth]{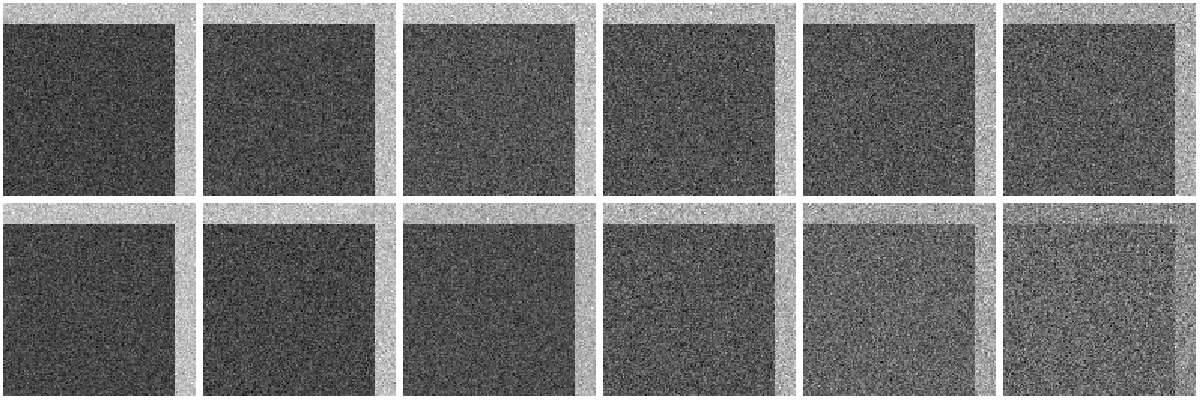}
    \caption{The top row shows the in-control images at time-points $t=$ $0$, $10$, $20$, $30$, $40$, and $50$ for Simulation 5. The bottom row shows the corresponding images for the out-of-control image sequence.}
    \label{oblm::sim4}
\end{figure}

\textbf{Simulation 5:} The last simulation differs from the previous ones as it demonstrates a scenario where the proposed method should not detect a change, despite an actual change occurring in the image sequence. In this case, the image consists of a stationary square with an intensity value of $0$ inside and $1$ outside. However, instead of any spatial movement, the intensities themselves change over time. The in-control sequence, the rate of change remains unaltered, while in out-of-control sequence, the rate of change alters after $t=20$. Note that the JLCs still remain unchanged in all images in both the sequences. Therefore, the proposed should not detect this type of change. Following similar procedure as in the previous simulation cases, we find that the detection delay to be approximately $20$ which is close to the in-control ARL value itself. This indicates that the algorithm behaves similarly to cases where no change is present. Such a property of the proposed algorithm is advantageous in certain real-world applications, such as ignoring variations due to shadows, changes in lighting, or time-of-the-day effects. In this case, $\kappa$ is selected to be $0.7$. The following functions represent the image sequences:
\begin{align*}
    &f_{IC}(x,y,t) = t*\alpha+(\beta -t\alpha)\mathbbm{1}\Big(max(x,y)>\frac{114}{128}\Big),\\
    &f_{OC}(x,y,t) = t*\alpha\big(1+\mathbbm{1}(t >20)\big)+(\beta -t\alpha)\mathbbm{1}\Big(max(x,y)>\frac{114}{128}\Big).
\end{align*}
Here $\beta$ is the starting intensity outside the square and $\alpha$ is the rate of change of intensity value. In this simulation, presented by Figure \ref{oblm::sim4}, we choose $\alpha = 0.005$, and $\beta = 1$.

\begin{figure}[ht!]
    \centering
    \includegraphics[width=12cm]{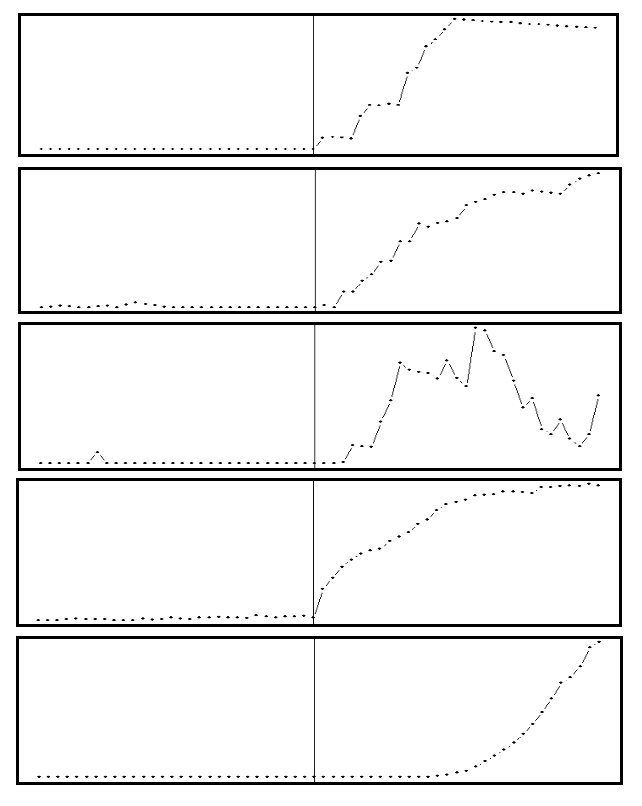}
\caption{CUSUM control charts corresponding to the five simulation scenarios. The panels are ordered from top to bottom, representing Simulation 1 through Simulation 5. Time is plotted along the $X$-axis, and the CUSUM charting statistic is plotted along the $Y$-axis.}
    \label{oblm::plot1}
\end{figure}

The following graphs show the CUSUM statistics over time for each of the scenarios. These graphs corresponds to one out-of-control image sequence. In Figure \ref{oblm::plot1}, time is plotted along the $X$-axis, the value of the CUSUM statistics are plotted along the $Y$-axis and the vertical line represents the actual change point, i.e., $t=20$. Note that in Simulation 5, the CUSUM chart starts to increase long after $t=20$. This is because when the difference between the intensity values inside and outside the square is getting smaller, the proposed method finds it increasingly more difficult to identify the JLCs.

\subsection{Comparison with state-of-the-art image monitoring methods}
To investigate the numerical performance of the proposed method, we consider the following state-of-the-art image monitoring methods as competing methods.

\textbf{(I) Supervised image monitoring algorithm by \cite{bui2018monitoring}:} It is an image monitoring and anomaly detection method in the framework of supervised learning. The central idea is to monitor the behavior of local residuals over time. To quantify deviations from the in-control state, \cite{bui2018monitoring} introduce a general spatial moving statistic (SMS) based on sample A-D (Anderson and Darling) statistic \cite{anderson1954test}. For online monitoring, the charting statistic at the $k$-th time-point $t_k$ is defined as: 
\begin{equation}
    S_{t_k}=\max_{(i,j)} \ \mbox{SMS}_{k,ij},
\end{equation}
where $\mbox{SMS}_{k,ij}$ denotes the spatial moving statistic at the $(i,j)$-th pixel at the $k$-th time-point. For a comprehensive discussion regarding this method, readers are referred to \cite{bui2018monitoring}. It is important to note that this method is a Shewhart type control chart. To implement the above procedure, we use \textit{spc4sts} package \cite{bui2021spc4sts} from CRAN-R.

\textbf{(II) Wavelet based image monitoring algorithm by \cite{koosha2017statistical}:} The wavelet-based method of image monitoring adopts a nonparametric profile monitoring framework where each image is decomposed into a set of 1-D profiles and monitored collectively using a GLR (generalized likelihood ratio) control chart. \cite{koosha2017statistical} refer the wavelet coefficients as frequency-domain features, and use a suitable wavelet transformation to extract these features from the images. In this current context, each row of an $n \times n$ image is treated as a profile, and extracting $d$ features per profile results in a wavelet coefficient vector of dimension $nd \times 1$ for each image.  Then, the central idea of online monitoring is to construct a GLR based control chart using the extracted $nd\times 1$ dimensional wavelet coefficient vectors.

We use the Average Run Length (ARL) and the standard deviations of the run lengths as measures of comparison of the concerned methods. For a test with $5\%$ significance level, the ARL should be $20$; and standard deviations of the run lengths should be close to $20$ as well. In this paper, we calculate these values empirically using $500$ independent replications per simulation data, for each method. These values are presented in the Table \ref{oblm:tab:sim_result}. Each cell in this table displays the estimated out-of-control ARL along with the corresponding standard deviation of the run lengths in parentheses.

\begin{table}[h]
\begin{center}
\begin{tabular}{c|c|c|c|c}
                                    &Type   & Proposed & Bui-Apley & Koosha-et.-al.\\
     \hline
     \multirow{2}{*}{Simulation 1}  &(IC)   & 21.586 (21.190) & 6.370 (0.499) & 18.614  (8.059) \\
                                    &(OC)   &  1.000 (0.000)  & 6.352 (0.478) &  14.166  (5.410)\\
     \hline
     \multirow{2}{*}{Simulation 2}  &(IC)   &21.062 (19.870) & 5.182 (12.728) & 25.000 (11.427) \\
                                    &(OC)   & 3.906 (1.190)& 4.078 (9.113)& 16.712  (6.381) \\
     \hline
     \multirow{2}{*}{Simulation 3}  &(IC)   & 21.508 (19.944)& 14.038 (14.265) &  19.064  (8.833)\\
                                    &(OC)   & 1.946 (0.860)& 13.262  (12.76)&  21.794 (24.768)\\
     \hline
     \multirow{2}{*}{Simulation 4}  &(IC)   &16.978 (22.200) & 21.026 (20.810) &  20.708 (9.208)\\
                                    &(OC)   & 1.000 (0.000)  & 13.656 (7.288) & 1.000 (0.000)  \\
      \hline
     \multirow{2}{*}{Simulation 5}  &(IC)   &22.304 (24.140) & 33.282 (5.252)&  25.000  (1.985)\\
                                    &(OC)   &18.998 (16.431) & 27.564 (3.577)&  15.368  (1.046)\\
     \hline
\end{tabular}
\caption{Phase II performance comparisons on various simulation settings.}\label{oblm:tab:sim_result}
\end{center}
\end{table}

From Table \ref{oblm:tab:sim_result}, we observe that for all simulated cases, both the in-control ARL and the standard deviation of the run length behave as expected. We also observe that in the first out-of-control scenario, the proposed method performs very well, much better than its competitors. However, in the second scenario, the detection delay appears relatively large. This is partly due to the nature of the simulated drift: the image sequence does not change between time points $20$ and $21$, but rather between $21$ and $22$, making the minimum possible detection delay equal to $2$. Additionally, the change in the drift pattern is very small, requiring slightly more than two time points for reliable detection. Similar behavior is observed in the third scenario. In the fourth scenario, we demonstrate that the proposed algorithm can detect a change in JLCs when there is no drift in the before the change. In the fourth scenario, the method by \cite{koosha2017statistical} perform similarly as well. Finally, in the fifth scenario, the in-control ARL, and the out-of-control ARL do not differ significantly, which aligns with the expected behavior, because JLCs do not change in this scenario.

The method proposed by \cite{bui2018monitoring} appears to detect changes prematurely in most in-control scenarios and it does not perform well in the out-of-control scenarios of Simulations 1-4. However, in the fifth simulation—where no drift is present—it behaves as expected. One of the major reason for this kind of performance of this method is attributed to the fact that this method is a Shewhart type control chart, and it is not designed to detect small incremental changes. However, this method can efficiently detect large pattern changes in images. For instance, pattern changes in the textured surfaces. The method developed by \cite{koosha2017statistical} exhibits similar limitations as the previous method, although it well as expected in the fourth scenario. Therefore, these types of image monitoring methods are often not practically useful in presence of small gradual changes.

\subsection{Real data application}
\begin{figure}[ht!]
    \centering
    \includegraphics[width=3cm,height=3cm]{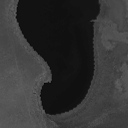}
    \includegraphics[width=3cm,height=3cm]{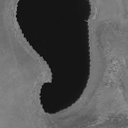}
    \includegraphics[width=3cm,height=3cm]{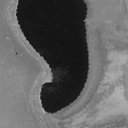}
    \includegraphics[width=3cm,height=3cm]{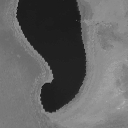}
    \caption{Sequence of the Aral sea images used for real data analysis. Images are captured in 2013, 2015, 2017, and, 2020. (Data  source: \href{https://earthexplorer.usgs.gov/}{U.S. Geological Survey.})}
    \label{oblm::fig:real_image}
\end{figure}

To illustrate the proposed ORT based image monitoring methodology, we consider a multi-temporal sequence of satellite images covering the southwestern region of the Aral Sea. Further details about the dataset can be found in Section \ref{oblm::data_description}. We analyze the temporal pattern of the shrinking of the Aral sea over a period of 2013 to 2024. Example images used for monitoring are presented in Figure \ref{oblm::fig:real_image}. 

\subsubsection{Data preparation}
\noindent In the real dataset, several scenes are compromised by heavy cloud cover and are therefore excluded from the analysis. Additionally, we also find missingness in the image data, certain stretches have no images at all, either due to unavailability of the original images or the removal of low-quality images. To minimize any impact on the proposed algorithm, we imputed these missing images by linearly interpolating the closest preceding and following images. We use the following equation to impute the missing images:
\[
\widehat{w}(x,y,t)=\frac{1}{(t_{i+1}-t_i)}\bigg[w(x,y,t_i)(t_{i+1}-t)+w(x,y,t_{i+1})(t-t_i)\bigg],\text{ where } t\in (t_i,t_{i+1}).
\]
By applying linear interpolation, we create an evenly spaced dataset covering the period from 2013 to 2024. We resize each image to the resolution $128 \times 128$. Subsequently, as a part of the pre-processing step, we perform image registration across the image dataset to remove any geometric misalignment by using \textit{rNiftyreg} \cite{image_registration} package from CRAN-R. For this analysis, we assume that the first $48$ images, i.e., those from 2013 to 2016 are in-control. We set the window width to $m_0 = 24$, meaning that preceding two years of data are used for prediction.

\subsubsection{Parameter selection}
\textbf{Allowance parameter $\kappa$:} To determine the value of the allowance parameter $\kappa$, we randomly select a few images from the first $48$ data-points of the prepared dataset, and impute the missing images, i.e., the images not selected in the sample, by the linear interpolation method. Using this bootstrapped sample, we calculate the values of the CUSUM statistic without the allowance parameter at each of the months $25$ through $48$. We then run the proposed algorithm on each bootstrapped dataset with allowance parameter set to 0, and compute the mean rate of change of the CUSUM statistic.

Suppose that $Q_{i,j}$ corresponds to month $j$ in $i$-th bootstrap sample and we have a total of $B$ bootstrap samples. We select $$\widehat{\kappa}=\frac{1}{B}\sum\limits_{i=1}^B\frac{1}{24}\sum\limits_{j=25}^{48}\frac{{Q}_{i,j}}{j}$$ as the value of the allowance parameter $\kappa$.

\textbf{Control limit $q_0$:} We use the above value of the allowance parameter to apply the proposed algorithm on the bootstrapped samples again. Next, we compute the sample mean and sample standard deviation of the CUSUM statistic, denoted by $\overline{Q}$ and $s_Q$, respectively. Finally, we use $\big(\overline{Q}+ \Phi^{-1}(0.95) s_Q \big)$ as the control limit $q_0$ for online drift pattern monitoring, where $\Phi^{-1}(0.95)$ is the $95$-th percentile of the standard normal distribution.

\subsubsection{Outcome of the analysis:}
\begin{figure}[ht!]
    \centering
    \includegraphics[width=\linewidth]{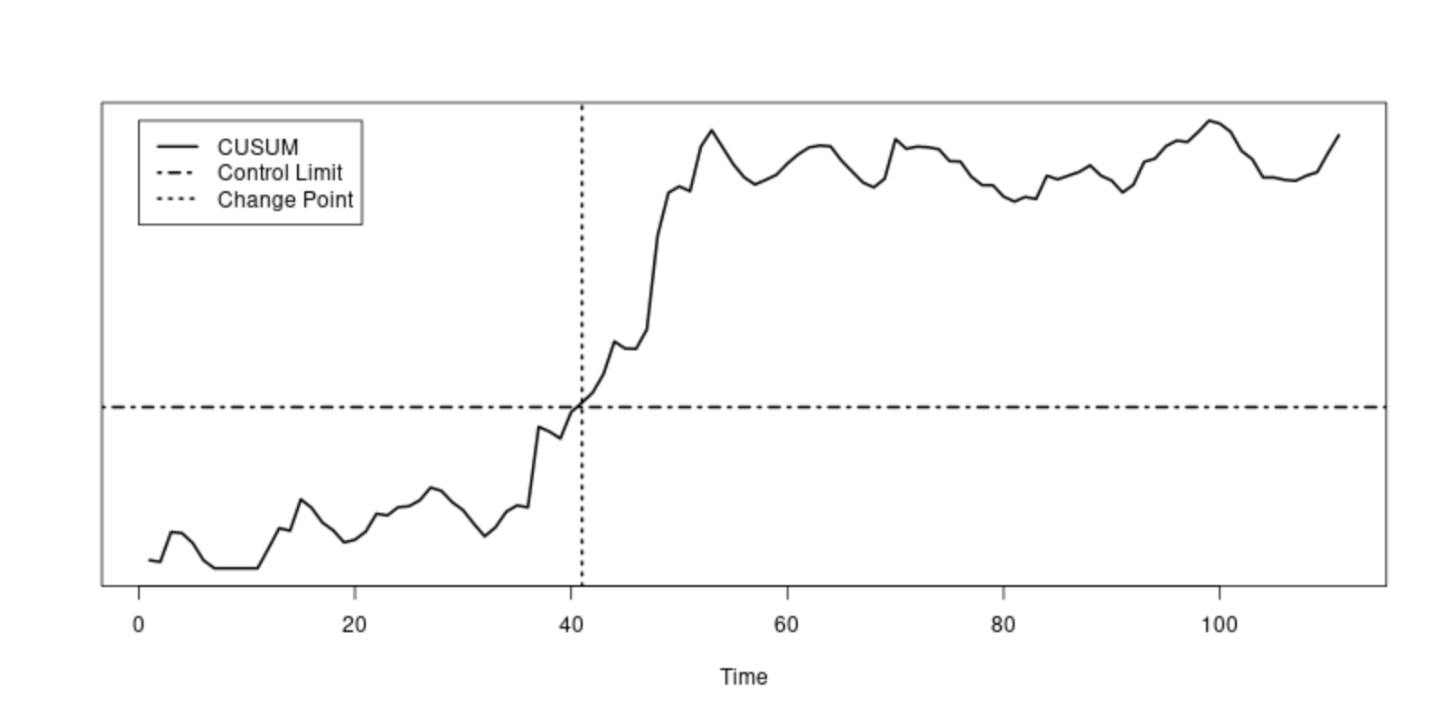}
    \caption{The proposed CUSUM chart for online drift pattern monitoring of the sequence of the Aral sea images. The time-point $0$ represents June of 2015, and one unit on the $X$-axis represents a month. The CUSUM charting statistic is plotted along the $Y$-axis.}
    \label{oblm::fig:real_online_monitoring}
\end{figure}

Using the procedure mentioned above, we construct the CUSUM chart displayed in Figure \ref{oblm::fig:real_online_monitoring}. In this chart, the time-point $0$ represents June of 2015, and one unit on the $X$-axis represents a month.. As the original data set start at June of 2013, the first estimate of CUSUM statistics that we can get is delayed by the window width $m_0 = 24$ used in this algorithm. Figure \ref{oblm::fig:real_online_monitoring} shows that there is a significant change in the drift pattern of the image sequence at the time-point $41$ representing November, 2018. Although there is hardly any recent study on the Aral sea shrinkage during 2013 to 2024, visual inspection suggests that a deceleration in the shrinking rate has started around the second half of 2018. Since shrinkage of the Aral sea is a global concern, various global initiatives have been taken to mitigate the problem. We believe that additional ecological studies are required to confirm the observed deceleration in the shrinkage rate of the Aral Sea.

\section{Concluding Remarks}\label{oblm::conclusion}
In this paper, we propose an efficient image monitoring scheme aimed at detecting shifts in the drift pattern of jump location curves (JLCs). A key innovation of the proposed approach lies in its decision tree-based framework for image surveillance, specifically employing oblique-axis regression trees (ORTs). While JLC-based image monitoring has shown promise in various practical applications, existing methods often fail to account for drift and have limited ability to accommodate gradual changes across sequential images. This research work contributes to the image monitoring literature by introducing a JLC-based methodology that effectively captures and tracks temporal evolution patterns in image sequences. The proposed method demonstrates superior performance across a variety of settings, as evidenced by comprehensive numerical studies that include both drift and non-drift scenarios in the in-control image sequence. Several promising avenues exist for extending this line of research. One immediate direction involves the development of drift monitoring techniques that can operate effectively in the presence of spatio-temporally correlated noise. Additionally, in real-world applications, image registration is often required as a pre-processing step. Thus, a monitoring framework that can inherently handle geometric misalignment of image objects across images would be highly desirable. Finally, the flexibility of the ORT-based framework suggests its potential applicability to a broader class of process monitoring problems beyond the scope of image data.

\bibliographystyle{apalike}
\typeout{}
\bibliography{bibliography.bib}

\newpage
\pagenumbering{arabic}

\begin{center}
{\large\bf Supplementary Materials of the paper titled ``Sequential Monitoring of Drift Patterns in Image Data''}
\end{center} 

\noindent {\bf S.1 \ A sketch of the proof of Theorem 1:}

We first introduce the following shorter notations to facilitate subsequent discussions.
\begin{itemize}
    \item $\mathcal{L}_n$ is a leaf node of the fitted tree when the resolution is $n^p$, $p$ being the dimension of the data. For 2-D image sequence over time, $p = 3$.
    \item $\mathcal{L}_n(x,y,t)$ is the leaf-node of the above tree which contains the point $(x,y,t)$.
    \item $\mu_k$ is the Lebesgue measure in $\mathbb{R}^k$.
\end{itemize}
The following results are provided by \cite{basak2025estimation}.
\begin{lemma}
If $\lim\limits_{n\rightarrow\infty}\mu_p(\mathcal{L}_n) \neq 0$, then $f$ is a.e. constant in $\bigcap\limits_{n}\mathcal{L}_n$.
\end{lemma}
\begin{lemma}
      If $\lim\limits_{n\rightarrow\infty}\mu_p(\mathcal{L}_n) = 0$ and $\bigcap\limits_n\mathcal{L}_n=\mathcal{L}$ lies on a $k$ dimensional affine subspace of $\mathbb{R}^p$ with $k<p$ and $\mu_k(\mathcal{L}) \neq 0$, then one of these is true:
     \begin{enumerate}[label = (\roman*)]
         \item $\mathcal{L}$ contains an entire JLC.
         \item $\mathcal{L}$ is contained entirely inside a JLC.
         \item $f$ is \it{a.e.} constant on $\mathcal{L}$. 
     \end{enumerate}\label{cor::kdim}
\end{lemma}
\begin{lemma}
     Let $\gamma(A)$ denote the maximum distance between two points in the set $A$. If $\lim\limits_{n \rightarrow \infty}\mu_p(\mathcal{L}_n) = 0$ and $\lim\limits_{n \rightarrow \infty}\gamma(\mathcal{L}_n)\neq 0$, then one of the following is true:
\begin{enumerate}[label=(\roman*)]
\item $\bigcap\limits_n\mathcal{L}_n$ contains an entire JLC.
\item $\bigcap\limits_n\mathcal{L}_n$ is contained entirely inside a JLC. 
\item $f$ is \it{a.e.} constant on $\bigcap\limits_n\mathcal{L}_n$.
\end{enumerate} 
\end{lemma}

\begin{lemma}
    Let $(x,y,t)$ be a non-singular continuity point of $f$. If $\lim\limits_{n\rightarrow\infty}\gamma(\mathcal{L}_n(x,y,t))=0$, then $\exists$ $N$ such that $\forall$ $m>N$, $\mathcal{L}_m(x,y,t)$ does not contain any discontinuity point. \label{oblm::dis}
\end{lemma}

\begin{lemma}
    $\lim\limits_{n\rightarrow \infty}\widehat{f}_n(x,y,t) =f(x,y,t)$ \textit{a.e.}, almost surely. 
\end{lemma}

\noindent \textit{Remarks:} These lemmas provide results only when the parameter space contains the time point $t$, i,e., $t$ is within the range of time points of images used in denoising. However, in the drift monitoring applications, that is not correct. In this algorithm, we predict the image observed at time $t_{k+1}$ based on the image data up to time point $t_k$ only. We now justify that below.

\noindent\textbf{Proof of Theorem 1:}
Under the given assumptions, $\widehat{f}_{n,t_{k}}(x,y,t)$ $\rightarrow$ $f_n(x,y,t)$ as $n \rightarrow \infty$, $\forall t\leq t_k$ as long as $(x,y,t)$ is not on a JLC. This is due to the results provided by \cite{basak2025estimation}.

Consider a point $(x,y,t_k)$ such that it is not on any JLC, i.e., $(x,y,t_k) \notin \bigcup_{\ell} \Big[\overline{\Gamma_\ell}\backslash int(\Gamma_\ell)\Big]$. Assume  that $(x,y,t_k) \in int(\Gamma_{\ell^\ast})$ for some $\ell^\ast$. As it is an interior point, $|t_{k+1}-t_k|\rightarrow0$, and under $\mathcal{H}_0$, the  drifts of the JLCs remain unchanged, we also have $(x,y,t_{k+1})\in int(\Gamma_{\ell^\ast})$.

Since $f$ is continuous at $(x,y,t_k)$, we have
\begin{equation}\tag{S.1}
    \Big|f(x,y,t_{k+1})-f(x,y,t_{k})\Big|\rightarrow 0.
    \label{oblm::tempeq1}
\end{equation}

As $(x,y,t_k)\notin \bigcup_{\ell} \Big[\overline{\Gamma_\ell}\backslash int(\Gamma_\ell)\Big]$, we have:
\begin{equation}\tag{S.2}
    \Big|f(x,y,t_k)-\widehat{f}_{t_k}(x,y,t_k)\Big|\rightarrow 0.
\end{equation}
The above result is due to \cite{basak2025estimation}.

Now, if $(x,y,t_{k+1})\in \mathcal{L}_k(x,y,t_k)$ then $\widehat{f}_{t_k}(x,y,t_k)=\widehat{f}_{t_k}(x,y,t_{k+1})$. Otherwise, we have one of the following two cases:

\noindent \textbf{Case 1}: $\gamma(\mathcal{L}(x,y,t_{k+1}))\rightarrow0$. Note that $\gamma(A)$ denotes the maximum distance between two points in the set $A$. In this case, we have
\begin{equation*}
    \Big|\widehat{f}_{t_k}(x,y,t_{k+1})-\widehat{f}_{t_k}(x,y,t_k)\Big|\leq C\Big|t_{k+1}-t_k+\gamma(\mathcal{L}(x,y,t_{k+1})\Big|\rightarrow 0,
\end{equation*}
as both $(x,y,t_k),(x,t,t_{k+1})\in \Gamma_i$.

\noindent \textbf{Case 2}:$\lim\limits_{n\rightarrow\infty}\mu_p(\mathcal{L}_n) \neq 0$ or $\gamma(\mathcal{L}(x,y,t_{k+1}))\not\rightarrow0$. Both of these scenarios imply that $f$ is constant in $\mathcal{L}(x,y,t_{k+1})$. Hence,
\begin{equation*}
    \Big|\widehat{f}_{t_k}(x,y,t_{k+1})-\widehat{f}_{t_k}(x,y,t_k)\Big|\leq C.\mbox{ dist}\big((x,y,t_k),\mathcal{L}(x,y,t_{k+1})\big)\leq C\Big|t_{k+1}-t_k\Big|\rightarrow 0.
\end{equation*}
Therefore, in both cases we have:
\begin{equation}\tag{S.3}
    \Big|\widehat{f}_{t_k}(x,y,t_{k+1})-\widehat{f}_{t_k}(x,y,t_k)\Big|\rightarrow0.
\end{equation}
From the Eqns. S.1, S.2, and S.3, we get $\delta_{t_{k+1}}(x,y)=\Big|f(x,y,t_{k+1})-\widehat{f}_{t_k}(x,y,t_{k+1})\Big|\rightarrow 0$.

Therefore, we have
\begin{align*}
    z_{t_k}(x,y) &=(w(x,y,t_k)-\widehat{f}_{n,t_{k}}(x,y,t_k))^2\\
                 &= (\varepsilon(x,y,t_k)+\delta(x,y,t_k))^2.
\end{align*}
Hence, $z_{t_k}(x,y) \overset{d}{=}\sigma^2 \chi^2_1 \bigg(\bigg[\frac{\delta_{t_k}(x,y)}{\sigma}\bigg]^2\bigg)$. Observe that $\frac{1}{n^2}\sum\limits_{(x,y)}z_{t_k}(x,y)$ has asymptotic mean $\sigma^2(1+E_{t_k})$, where $E_{t_k}=\frac{1}{\sigma^2}\int\limits_0^1\int\limits_0^1\delta_{t_k}(x,y)^2dxdy$. Noe that $E_{t_k}\rightarrow0$ as $\delta_{t_K}\rightarrow0$ {\it a.e.}, and it is bounded by $2$.
Hence, we have
\begin{align*}
    \theta^2\Lambda_{t_k} & \overset{d}{=} \frac{\sigma^2\chi^2_{(n^2-K(\mathcal{I}_{t_k}))}(\frac{1}{\sigma^2}\sum\delta_{t_k}(x,y)^2)}{(n^2-K(\mathcal{I}_{t_k}))}
     \approx \frac{\sigma^2\chi^2_{n^2}(\frac{1}{\sigma^2}\sum\delta_{t_k}(x,y)^2)}{n^2}.\\
    & \implies n\left(\frac{\theta^2}{\sigma^2}\Lambda_{t_k}-1-E_{t_k}\right) \overset{d}{=}\mathcal{N}\left(0,2+4E_{t_k}\right).\\
    & \implies \frac{n}{\sqrt{2}}\left(\Lambda_{t_k}-(1+E_{t_k})\frac{\sigma^2}{\theta^2}\right) \overset{d}{=} \mathcal{N}\left(0,
    \frac{\sigma^2}{\theta^2}(1+2E_{t_k})\right).
\end{align*}
Since we consider $\theta^2=\sigma^2(1+E_{t_k})$, we have \begin{equation}
\label{oblm::h0dist}\tag{S.4}
    \frac{n}{\sqrt{2}}(\Lambda_{t_k}-1)\overset{d}{=} \mathcal{N}\left(0,1+\frac{E_{t_k}}{1+E_{t_k}}\right) \overset{d}{\approx}\mathcal{N}(0,1)  \text{ as } E_{t_k}\rightarrow 0.
\end{equation}

Under $\mathcal{H}_1$, at least one JLC has not continued to the new time-point. Hence the  estimates will be incorrect at the points which were near the previous JLC. This deviation should be proportional to the jump size, say $\Delta$. Note that in this context, we refer to the JLCs in the 3-D space, not in individual 2-D images. Since any linear JLC has approximately $O({n})$ many pixels, we can conclude that
\[
\theta^2\Lambda_{t_k} \succ  \frac{1}{n^2}\sigma^2\chi^2_{n^2}\left(\frac{1}{\sigma^2}\sum\delta_{t_k}(x,y)^2+O(n)\Delta^2\right).
\]
Then, proceeding similarly as displayed in Eqn. \eqref{oblm::h0dist}, we have
\[
\frac{n}{\sqrt{2}}\Big(\Lambda_{t_k}-1-O(1)\Delta^2\Big) \succ \mathcal{N}\left(0,1+\frac{E_{t_k}+O(1)\Delta^2}{1+E_{t_k}}\right). 
\]

\end{document}